\documentclass[cameraready]{Interspeech}
\title{\textbf{DTT-BSR+: A Generative-Regression Cascade for Music Source Restoration}}

\author[affiliation={1}, equalcontribution]{Youran}{Ni}
\author[affiliation={1}, equalcontribution]{Shihong}{Tan}
\author[affiliation={2}]{Yuzhu}{Wang}
\author[affiliation={1}, correspondingauthor]{Gongping}{Huang}

\address{
    $^1$ School of Electronic Information, Wuhan University, China \\
    $^2$ Signal Processing Research Center, Tampere University, Tampere, Finland
}

\email{2023302121082@whu.edu.cn, shihongtan@whu.edu.cn,yuzhu.wang@tuni.fi,gongpinghuang@whu.edu.cn}


\definecolor{red_cool}{rgb}{0.5, 0.0, 0.0}
\usepackage{comment}
\usepackage{multirow}
\usepackage{graphicx}
\usepackage{booktabs}
\usepackage{amsmath}
\usepackage{siunitx}

\begin{document}

\maketitle

% Abstract
\begin{abstract}

Music source restoration (MSR) requires jointly addressing source unmixing and the inversion of non-linear production effects. Current methods struggle to achieve accurate target signal reconstruction while maintaining semantic consistency. To address this limitation, we propose DTT-BSR+, a two-stage cascade MSR system that decouples distribution fitting from signal reconstruction into separate stages. A generative DTT-BSR separator in the first stage produces stems matching the prior of clean sources, and a modified Demucs network in the second stage enhances the first stage output using time-domain and multi-resolution spectral losses. DTT-BSR+ improves multi-mel signal-to-noise ratio (MMSNR) over the single-stage DTT-BSR across all stems, and surpasses the state-of-the-art X-LANCE MSR system on five stems. We also reveal through Fréchet Audio Distance (FAD) decomposition an implicit trade-off between signal reconstruction accuracy and semantic distribution fitting across stems.
\end{abstract}

\section{Introduction}
 \label{sec:intro}
 Music source restoration (MSR)~\cite{zang2025msr} aims to estimate clean, unprocessed stems from a mixture. The difficulty lies in the non-linear production effects applied to each stem during music creation, such as dynamic range compression~\cite{compressor, mcnally1984dynamic} and codec encoding~\cite{Sripada2006MP3DI}. Accurate estimation requires jointly addressing source unmixing and the inversion of production effects, going beyond the music source separation (MSS) task~\cite{10096956,wang2024melroformervocalseparationvocal,wu2025lowlatency} which models the mixture as a sum of sources without accounting for such effects~\cite{cano2019musical}. In practice, MSR can be applied to a wide range of music production workflows, such as stem recovery and remixing from finished mixes.

MSR research remains limited, with the most recent advances primarily emerging from the ICASSP 2026 MSR Challenge~\cite{zang2026summaryinauguralmusicsource}. Existing methods can be broadly categorized into two groups. The first adopts a multi-stage design, where each stage is trained independently with its own objective. The X-LANCE MSR system~\cite{zhu2026sjtuxlancelabmsr} applies three pretrained BSRoformer~\cite{10446843} modules sequentially for separation, dereverberation, and denoising, achieving state-of-the-art performance on both objective and subjective evaluations. CUPAudioGroup system~\cite{zang2026summaryinauguralmusicsource} ensembles three complementary MSS models initialized from pretrained weights, leveraging model diversity to improve restoration quality. The second group builds joint systems optimized under multiple losses simultaneously. DTT-BSR~\cite{tan2026dttbsrganbaseddttnetrope} combines a dual-path TFC-TDF U-Net~\cite{10448020} with a RoPE transformer bottleneck, trained with a joint GAN and reconstruction loss. Hachimi system~\cite{zang2026summaryinauguralmusicsource} similarly adopts a mel-band backbone with combined GAN and reconstruction losses. While existing methods achieve reasonable Fr\'{e}chet Audio Distance (FAD)~\cite{kilgour19_interspeech} scores, multi-mel signal-to-noise ratio (MMSNR) remains limited across all systems, suggesting that the restored stem matches the semantic distribution of the target but signal reconstruction remains inaccurate.

\begin{figure}[t]
  \centering
  \includegraphics[width=0.95\linewidth]{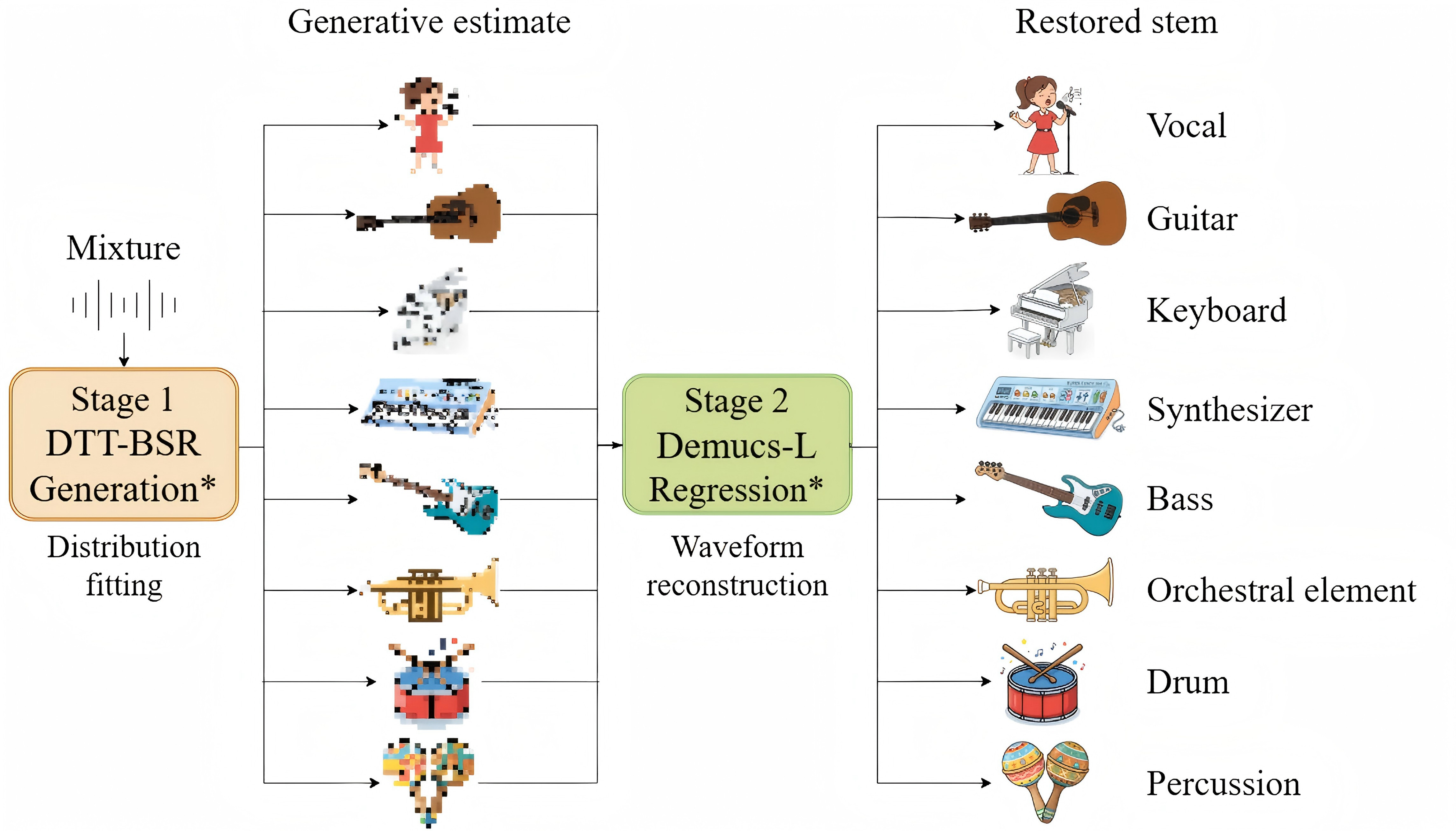}
  % \vspace{-1ex}
  \caption{Overview of the proposed DTT-BSR+ system.}
  \label{fig:model_high_level}
  \vspace{-3ex}
\end{figure}

To address this limitation, we propose DTT-BSR+, a two-stage cascade MSR system as shown in Figure~\ref{fig:model_high_level}. This cascade design is similar to the X-LANCE system. The difference is that the X-LANCE decomposes MSR into three sub-stages of separation, dereverberation, and denoising, whereas DTT-BSR+ decomposes it into a semantic distribution fitting stage and a signal reconstruction stage. The first stage uses DTT-BSR as a generative separator to estimate stems that match the prior distribution of clean sources. The second stage employs a modified Demucs network~\cite{defossez2019music} trained to minimize the difference between its output and the ground-truth clean stem, using the first stage output as input, optimized with time-domain and multi-resolution spectral losses to reconstruct the target signal.
\begin{figure*}[t]
  \centering
  \includegraphics[width=1\linewidth]{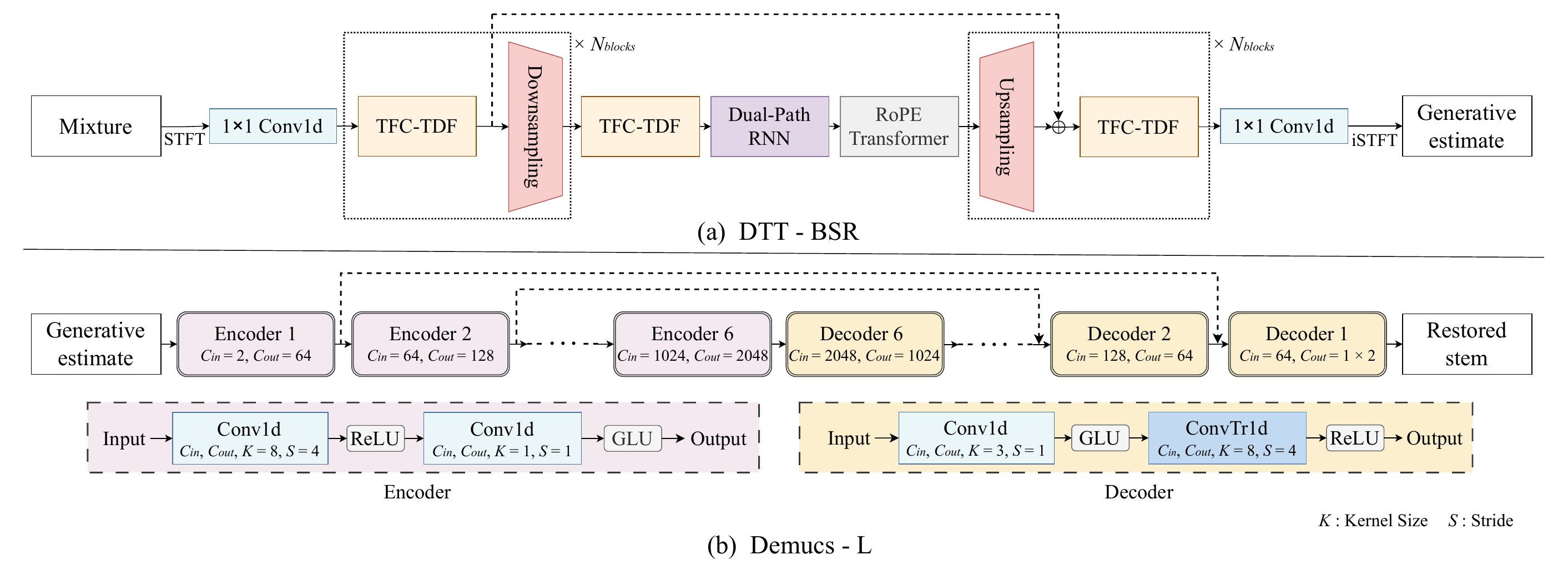}
  \vspace{-3ex}
  \caption{Architecture of the proposed DTT-BSR+ system. (a) DTT-BSR serves as the first stage, producing a generative estimate from the degraded mixture. (b) Demucs-L serves as the second stage, reconstructing the target stem from the first stage output.}
  \label{fig:model}
  \vspace{-2ex}
\end{figure*}
The main contributions of this paper are summarized as follows. (1) We propose DTT-BSR+, a novel two-stage cascade MSR system. Experimental results show that DTT-BSR+ improves MMSNR over the single-stage DTT-BSR across all eight stems, and surpasses the state-of-the-art X-LANCE MSR system on five stems including Vocals, Guitars, Synthesizers, Bass, and Drums. (2) Through ablation experiments, we validate the design choices of the proposed DTT-BSR+, and identify its limitations for Percussions, where severe signal distortion in the first stage cannot be restored by the second stage, and a single-stage MSG network~\cite{schaffer2022musicseparationenhancementgenerative} is shown to be effective for this stem. (3) We show that improving signal reconstruction accuracy does not consistently improve FAD across stems, suggesting an implicit trade-off between accurate signal reconstruction and semantic distribution fitting. FAD decomposition further reveals that FAD degradation is primarily driven by semantic mean shift rather than changes in distribution variance.

\section{Problem Formulation}
\label{sec:problem}
In MSS, the time-domain observed mixture $y \in \mathbb{R}^{T}$ is modeled as a sum of $C$ source signals, i.e., $y = \sum_{c=1}^{C} x_c$, where $x_c \in \mathbb{R}^{T}$ denotes the $c$-th source, and the goal is to estimate each $x_c$ from $y$~\cite{cano2019musical}. MSR extends this by modeling each source as $x_c = f_c(s_c)$, where $s_c \in \mathbb{R}^{T}$ is the clean, unprocessed stem and $f_c$ models the production effects, which may include spectral filtering~\cite{9831033}, dynamic range compression, harmonic distortion~\cite{schuck2016audio}, reverberation\cite{smith2010physical}, and codec degradation. The MSR mixture model thus becomes
\begin{align}
    y = \sum_{c=1}^{C} f_c(s_c),
    \label{eq:msr}
\end{align}
and the goal is to estimate each clean stem $s_c$ from $y$~\cite{zang2025msr}. In this paper, we consider $C = 8$ stems: Vocals, Guitars, Keyboards, Synthesizers, Bass, Orchestral elements, Drums, and Percussions.
The formulation reveals that MSR poses a greater challenge than MSS. Estimating each $s_c$ from $y$ requires simultaneously addressing the mixing of sources and the non-linear, irreversible nature of the effect functions $f_c$, both of which are coupled in the observed mixture. Since $f_c$ is many-to-one, the estimation of $s_c$ is inherently ill-posed.
% To define the MSR problem, We firstly define $y \in \mathbb{R}^{C \times T}$ as the observed waveform, where $C$ denotes the channels of the waveform and $T$ denotes the  sample points. We model $y$ to be the combination of a set of degraded signals, which can be stated as

% \begin{align}
% 	y=\sum_{i=1}^{n}\mathcal{D}_i(s_i)
% 	\label{eq:pf}
% \end{align}

% where $s_i$ denotes the clean, unprocessed stem and $\mathcal{D}_i \in \mathcal{D}$ denotes degradation functions and degradation set. In this paper, we discuss $n=8$ stems: Vocals, Guitars, Keyboards, Synthesizers, Bass, Orchestral elements, drums, and percussions. The degradation set $\mathcal{D}$ discussed in this paper is mentioned in section \ref{sec:intro}.

% Hence, the MSR problem aims to obtain $s_i$ from $y$, given the observed wave $y$, the possible degradation set $\mathcal{D}$, and the priors $p_i$, denotes distributions of $s_i$, with which each $s_i$ is consistent. We aim to generate $s_i$ that is plausible and neutral, which means each recovered $s_i$ can satisfy equation (\ref{eq:pf}) and belongs to $p_i$~\cite{zang2025msr}.

\section{Proposed Method: DTT-BSR+}
\label{sec:model}
We propose DTT-BSR+, a two-stage cascade MSR system illustrated in Figure~\ref{fig:model}. The first stage employs a GAN-based separator to separate the target stem from the degraded mixture, with the adversarial objective constraining the output distribution toward that of the clean and unprocessed stem. The second stage enhances the first stage output using a regression-based network to improve waveform-level reconstruction accuracy.
\subsection{First stage: GAN-based Separation}

The first stage adopts DTT-BSR~\cite{tan2026dttbsrganbaseddttnetrope} as the separation backbone. Given the mixture $y$, DTT-BSR produces an estimate $\hat{s}_{\text{target}}$ of the target clean stem $s_{\text{target}}$. The primary objective of this stage is to obtain a stem estimate whose distribution is consistent with the prior distribution of the clean target stem. DTT-BSR is trained with a joint objective combining adversarial and reconstruction losses. The reconstruction loss within the joint objective constrains the generation space, preventing the generated output from deviating excessively from the target in the time-frequency domain. As a result, $\hat{s}_{\text{target}}$ achieves reasonable semantic consistency but exhibits limited waveform-level accuracy~\cite{tan2026dttbsrganbaseddttnetrope}. These limitations motivate the second stage, which takes $\hat{s}_{\text{target}}$ as input and optimizes for waveform-level reconstruction.
\subsection{Second stage: Waveform-level Reconstruction}

The second stage takes $\hat{s}_{\text{target}}$ as input and produces $\hat{s}_{\text{final}}$ as the reconstructed estimate of the target clean stem $s_{\text{target}}$. We adopt a modified Demucs~\cite{defossez2019music} architecture referred to as Demucs-L as the reconstruction network. Demucs-L is based on a 1D convolutional U-Net with Gated Linear Units (GLU)~\cite{dauphin2017language} and strided convolutions. To prioritize local modeling over global sequence modeling, we remove the bidirectional long short-term memory (BLSTM) bottleneck present in the standard Demucs architecture. This modification constrains the temporal receptive field, ensuring that the second stage focuses on local waveform-level reconstruction without altering the output distribution established by the first stage.
\subsection{Optimization Objective}
We train Demucs-L with a regression-based objective combining a time-domain $L_1$ loss and a multi-resolution short-time Fourier transform (MR-STFT) loss~\cite{yamamoto2020parallel} to jointly constrain reconstruction accuracy in both the time and time-frequency domains,
\begin{align}
    \mathcal{L}_{\text{Total}} = \lambda_1 \mathcal{L}_{L1} + \lambda_2 \mathcal{L}_{\text{MR-STFT}},
    \label{eq:total_loss}
\end{align}
where $\lambda_1$ and $\lambda_2$ are weighting coefficients. The $L_1$ loss $\mathcal{L}_{L1} = \lVert s_{\text{target}} - \hat{s}_{\text{final}} \rVert_1$ enforces point-wise waveform alignment between the reconstructed output and the ground-truth clean stem. The MR-STFT loss $\mathcal{L}_{\text{MR-STFT}}$ averages spectral reconstruction errors across $M$ STFT resolutions with varying FFT sizes, hop lengths, and window sizes, capturing reconstruction accuracy at multiple time-frequency scales. The two loss terms are complementary, with $\mathcal{L}_{L1}$ constraining temporal fidelity and $\mathcal{L}_{\text{MR-STFT}}$ constraining spectral fidelity across multiple resolutions.

\begin{table*}[t]
    \centering
    \footnotesize
    \caption{Per-stem results on MSRBench, evaluated with MMSNR ($\uparrow$), Zimtohrli ($\downarrow$), and FAD-CLAP ($\downarrow$).}
    \vspace{-1.5ex}
    \label{table:main_results}
    \setlength{\tabcolsep}{8pt} % 稍微缩小间距以容纳新加的 Avg 列
    \begin{tabular}{llccccccccc}
        \toprule
        \textbf{Method} & \textbf{Metric} & \textbf{Voc} & \textbf{Gtr} & \textbf{Key} & \textbf{Syn} & \textbf{Bass} & \textbf{Orc} & \textbf{Drm} & \textbf{Prc} & \textbf{Avg.} \\
        \midrule
        \multirow{3}{*}{BSRNN (Baseline)} 
        & MMSNR($\uparrow$)       & 3.24 & 1.15 & 0.57 & 0.35 & 2.37 & 0.34 & 1.83 & 0.01 & 1.23 \\
        & Zimtohrli($\downarrow$) & 0.020 & 0.019 & 0.021 & 0.024 & 0.019 & 0.027 & 0.022 & 0.023 & 0.022 \\
        & FAD-CLAP($\downarrow$)  & 0.352 & 0.476 & 1.032 & 0.911 & 0.668 & 0.758 & 0.682 & 1.294 & 0.772 \\
        \midrule
        \multirow{3}{*}{X-LANCE-MSR} 
        & MMSNR($\uparrow$)       & 3.36 & 1.99 & \textbf{2.76} & 1.22 & 4.22 & \textbf{1.59} & 2.48 & \textbf{0.59} & 2.28 \\
        & Zimtohrli($\downarrow$) & \textbf{0.019} & 0.019 & 0.017 & 0.026 & 0.014 & 0.023 & 0.020 & 0.027 & 0.021 \\ 
        & FAD-CLAP($\downarrow$)  & 0.353 & 0.398 & \textbf{0.523} & \textbf{0.639} & \textbf{0.439} & 0.656 & 0.533 & \textbf{0.973} & \textbf{0.564} \\
        \midrule
        \multirow{3}{*}{DTT-BSR} 
        & MMSNR($\uparrow$)       & 3.34 & 1.12 & 0.93 & 0.44 & 2.49 & 0.39 & 2.24 & 0.07 & 1.38 \\
        & Zimtohrli($\downarrow$) & \textbf{0.019} & 0.019 & 0.020 & 0.023 & 0.016 & 0.029 & 0.020 & 0.028 & 0.022 \\
        & FAD-CLAP($\downarrow$)  & 0.298 & \textbf{0.367} & 0.705 & 0.836 & 0.603 & \textbf{0.576} & 0.434 & 1.126 & 0.618 \\
        \midrule
        \multirow{3}{*}{\textbf{DTT-BSR +}} 
        & MMSNR($\uparrow$)       & \textbf{6.72} & \textbf{3.51} & 2.19 & \textbf{2.30} & \textbf{9.29} & 1.57 & \textbf{8.79} & 0.42  & \textbf{4.35} \\ 
        & Zimtohrli($\downarrow$) & \textbf{0.019} & \textbf{0.016} & \textbf{0.013} & \textbf{0.016} & \textbf{0.010} & \textbf{0.017} & \textbf{0.012} & \textbf{0.014} & \textbf{0.015} \\ 
        & FAD-CLAP($\downarrow$)  & \textbf{0.289} & 0.640 & 1.112 & 0.959 & 0.546 & 0.980 & \textbf{0.395} & 1.016 & 0.742 \\ 
        \bottomrule
    \end{tabular}
    \vspace{-2ex}
\end{table*}
\vspace{-1ex}
\section{Experiment}
\vspace{-1ex}
\subsection{Dataset}
The proposed system is trained and evaluated on MSRBench~\cite{zang2025msrbenchbenchmarkingdatasetmusic}, a benchmark dataset for music source restoration comprising $3250$ stereo clips at $48$~kHz, each $10$~seconds in duration, with parallel degraded mixtures and unprocessed ground-truth stems. The dataset is split into $2600$ training, $325$ validation, and $325$ test clips. The first stage DTT-BSR is pre-trained on RawStems~\cite{zang2025msr} following ~\cite{tan2026dttbsrganbaseddttnetrope}, a dataset disjoint from MSRBench. To train the second stage, we run inference with the frozen first stage on the 2600 MSRBench training clips to generate an intermediate dataset of 2600 input-target pairs for each stem, where the input is the first stage output and the target is the corresponding ground-truth stem. The $325$ test clips are reserved for evaluation and are not used in any stage of training.
\vspace{-1ex}
\subsection{Training Configurations}
% \yuzhu{Q: What is the batch size for the second stage training? \\
% Q: What is the duration for each batch? Do we use chunk operation during training? \\
% Q: Do we use data augmentation methods? \\}
The first stage DTT-BSR is trained following~\cite{tan2026dttbsrganbaseddttnetrope}. The second stage Demucs-L uses a $6$-layer encoder and a symmetric $6$-layer decoder with skip connections, where the convolutional layers have a kernel size of $8$ and a stride of $4$, starting with $64$ channels that double at each downsampling step. We train Demucs-L using the Adam optimizer~\cite{kingma2015adam} with a learning rate of $2 \times 10^{-4}$ for $150$ epochs on an NVIDIA RTX 4090 GPU, with a batch size of $16$. The loss coefficients are set to $\lambda_1 = 10.0$ and $\lambda_2 = 1.0$, and the checkpoint with the lowest validation loss is retained as the final model. All hyperparameters were determined through pre-experiments on the validation set. 

Two data augmentation strategies are applied during training. A random phase offset is applied to the input in the frequency domain to improve robustness to phase variations. Additionally, the first stage output is replaced by the ground-truth target stem with a probability of $10$~\%, improving generalization and preventing the network from overfitting to the first stage output distribution. The $10$-second audio clips are chunked into $1$-second segments during training and evaluated without chunking during inference.
\vspace{-1ex}
\subsection{Evaluation Metrics}
We assess restoration quality using three objective metrics. First, we calculate the scale-invariant MMSNR to measure spectral fidelity. MMSNR averages the signal-to-noise ratio across $N$ Mel-spectrogram resolutions to capture reconstruction accuracy at multiple time-frequency scales,
\begin{align}    
\mathrm{MMSNR} = \frac{1}{N}\sum_{i=1}^{N}  \frac{\sum_{T,F} \mathrm{Mel}_{r_i}(T,F)^2}{\sum_{T,F} (\mathrm{Mel}_{r_i}(T,F) - \alpha\hat{\mathrm{Mel}}_{r_i}(T,F))^2} ,
\end{align}
where $r_i$ denotes the $i$-th FFT resolution, $\mathrm{Mel}_{r_i}$ represents the corresponding Mel-spectrogram transformation, and $\alpha$ is the optimal scaling factor computed in the time domain as $\alpha = \frac{\langle s_{\text{target}}, \hat{s}_{\text{final}} \rangle}{\langle \hat{s}_{\text{final}}, \hat{s}_{\text{final}} \rangle}$ to ensure scale-invariance.
% \lVert s_{\text{target}} - \hat{s}_{\text{final}}
\begin{figure}[t]
    % Group 1: Bass (Left Column)
    \centering
    \includegraphics[width=0.85\linewidth]{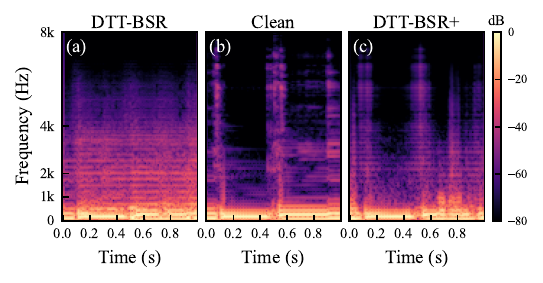} 
    \vspace{-3ex}
    \caption{Mel spectrograms of the Bass stem: (a) DTT-BSR output, (b) clean stem, and (c) DTT-BSR+ output.}
    \label{fig:bass_spec}
    \vspace{-3ex}
\end{figure}

Second, we employ Zimtohrli~\cite{alakuijala2025zimtohrli} to measure perceptual similarity. This psychoacoustic metric simulates the human auditory system using a Gammatone filterbank and non-linear loudness normalization, capturing perceptual distortions that are not directly reflected in signal-level reconstruction metrics such as MMSNR.

Finally, we evaluate semantic consistency using the Fr\'{e}chet Audio Distance based on CLAP embeddings (FAD-CLAP)~\cite{kilgour19_interspeech},
\begin{align}    
\mathrm{FAD\text{-}CLAP} = \underbrace{\lVert \mu_r - \mu_g \rVert_2^2}_{\mathcal{D}_\mu (\text{mean term})} + \underbrace{\mathrm{Tr}(\Sigma_r + \Sigma_g - 2\sqrt{\Sigma_r\Sigma_g})}_{\mathcal{D}_\Sigma (\text{covariance term})},
\end{align}
where $(\mu_r, \Sigma_r)$ and $(\mu_g, \Sigma_g)$ denote the mean and covariance matrices of the CLAP embeddings of the reference and restored audio, respectively. FAD-CLAP measures the distributional distance between the reference and restored audio in the CLAP embedding space, where the mean term $\mathcal{D}_\mu$ captures the shift in the semantic center and the covariance term $\mathcal{D}_\Sigma$ reflects the difference in distribution diversity.
\vspace{-1ex}
\section{Results and Analysis}
\label{sec:ablation_arch}
We evaluate DTT-BSR+ against three systems. BSRNN~\cite{10121418} is a single-stage system serving as the official baseline of the ICASSP 2026 MSR Challenge. X-LANCE-MSR is a multi-stage system and the challenge winner representing the current state-of-the-art. DTT-BSR is a competitive single-stage system that also serves as the first stage component of DTT-BSR+.
\vspace{-1ex}
\subsection{Performance Comparison}

Table~\ref{table:main_results} presents the per-stem results of DTT-BSR+ alongside all three comparison systems. DTT-BSR+ achieves consistent MMSNR improvements over DTT-BSR across all eight stems, demonstrating that the second stage effectively enhances the first stage output. The largest gains are observed on Bass ($2.49$~dB to $9.29$~dB), Drums ($2.24$~dB to $8.79$~dB), and Vocals ($3.34$~dB to $6.72$~dB). Figure~\ref{fig:bass_spec} shows that the DTT-BSR+ output exhibits time-frequency characteristics closer to the target stem compared to the DTT-BSR output on Bass. Compared to X-LANCE-MSR, DTT-BSR+ achieves higher MMSNR on Vocals, Guitars, Synthesizers, Bass, and Drums, with the largest margins on Bass ($9.29$~dB vs $4.22$~dB) and Drums ($8.79$~dB vs $2.48$~dB), while X-LANCE-MSR retains an advantage on Keyboards and Orchestral. Across all systems, Percussions remains the most challenging stem, with the best MMSNR among all methods reaching only $0.59$~dB. DTT-BSR+ achieves the best Zimtohrli score across all eight stems, indicating that the gains in signal reconstruction accuracy are consistent with improvements in perceptual quality. Regarding FAD-CLAP, a comparison between DTT-BSR and DTT-BSR+ reveals two distinct trends across stems. On Vocals, Bass, Drums, and Percussions, FAD-CLAP decreases alongside the MMSNR improvement, indicating that both signal fidelity and semantic consistency benefit from the second stage. On Guitars, Keyboards, Synthesizers, and Orchestral, FAD-CLAP increases despite higher MMSNR, revealing a divergence between signal reconstruction accuracy and semantic consistency on these stems.
% \begin{table}[tb]
%     \centering
%     \footnotesize
%     \caption{Ablation study on MSRBench (MMSNR $\uparrow$).}
%     \vspace{-1ex}
%     \label{table:ablation_arch}
%     \setlength{\tabcolsep}{2pt}
%     \begin{tabular}{ccccccccccc}
%     \toprule
%     \textbf{\#} & \textbf{Stage-1} & \textbf{Stage-2} & \textbf{Voc} & \textbf{Gtr} & \textbf{Key} & \textbf{Syn} & \textbf{Bass} & \textbf{Orc} & \textbf{Drm} & \textbf{Prc} \\
%     \midrule
%     1 & DTT-BSR & ---       & 3.34 & 1.12 & 0.93 & 0.44 & 2.49 & 0.39 & 2.24 & 0.07 \\
%     2 & ---      & MSG      & 4.70 & 0.07 & 0.30 & 0.34 & 7.33 & 0.14 & 7.89 & \textbf{2.47} \\
%     3 & DTT-BSR & MSG      & 5.70 & 3.21 & 0.90 & 0.29 & 8.64 & 0.08 & \textbf{9.01} & 0.13 \\
%     \midrule
%     4 & DTT-BSR & Demucs-L & \textbf{6.72} & \textbf{3.51} & \textbf{2.19} & \textbf{2.30} & \textbf{9.29} & \textbf{1.57} & 8.79 & 0.42 \\
%     \bottomrule
%     \end{tabular}
%     \vspace{-2ex}
% \end{table}
% \vspace{-1ex}
\begin{table}[tb]
    \centering
    \footnotesize
    \caption{Ablation study on MSRBench (MMSNR $\uparrow$).}
    \vspace{-1ex}
    \label{table:ablation_arch}
    \setlength{\tabcolsep}{1.5pt}
    \begin{tabular}{cccccccccccc}
    \toprule
    \textbf{Stage-1} & \textbf{Stage-2} & \textbf{Voc} & \textbf{Gtr} & \textbf{Key} & \textbf{Syn} & \textbf{Bass} & \textbf{Orc} & \textbf{Drm} & \textbf{Prc} & \textbf{Avg.} \\
    \midrule
    DTT-BSR & ---       & 3.34 & 1.12 & 0.93 & 0.44 & 2.49 & 0.39 & 2.24 & 0.07 & 1.38 \\
    ---      & MSG      & 4.70 & 0.07 & 0.30 & 0.34 & 7.33 & 0.14 & 7.89 & \textbf{2.47} & 2.91 \\
    DTT-BSR & MSG      & 5.70 & 3.21 & 0.90 & 0.29 & 8.64 & 0.08 & \textbf{9.01} & 0.13 & 3.50 \\
    \midrule
    DTT-BSR & Demucs-L & \textbf{6.72} & \textbf{3.51} & \textbf{2.19} & \textbf{2.30} & \textbf{9.29} & \textbf{1.57} & 8.79 & 0.42 & \textbf{4.35} \\
    \bottomrule
    \end{tabular}
    \vspace{-2ex}
\end{table}
\vspace{-1ex}
\subsection{Ablation Study}

Table~\ref{table:ablation_arch} presents the architectural ablation results, where Demucs-L denotes the modified Demucs network with the BLSTM bottleneck removed, restricting the temporal receptive field to focus on local waveform-level reconstruction rather than global sequence modeling. We also include MSG, a widely used GAN-based music enhancement network, as an alternative second stage to compare generative and regression-based reconstruction approaches. Comparing rows $1$ and $4$, adding the second stage consistently improves MMSNR across all eight stems, confirming that Demucs-L effectively enhances the first stage output. Comparing rows $3$ and $4$, replacing MSG with Demucs-L as the second stage yields higher MMSNR on six stems, with large improvements on Keyboards ($0.90$~dB to $2.19$~dB) and Synthesizers ($0.29$~dB to $2.30$~dB), while MSG retains an advantage on Drums ($9.01$~dB vs $8.79$~dB). Comparing the single-stage systems (rows $1$ and $2$) against the cascade systems (rows $3$ and $4$), the cascade configurations achieve higher MMSNR than their single-stage counterparts on most stems, demonstrating the overall advantage of the multi-stage design. The exception is Percussions, where the single-stage MSG achieves the best MMSNR among all configurations ($2.47$~dB), while both cascade systems degrade substantially, with rows $3$ and $4$ reaching $0.13$~dB and $0.42$~dB respectively, indicating that the first stage processing introduces a distortion on this stem that cannot be improved by the second stage.
\vspace{-1ex}
\subsection{FAD-CLAP Analysis}
\label{sec:fad_tradeoff}

\begin{figure}[t]
    \centering
    \includegraphics[width=0.8\linewidth]{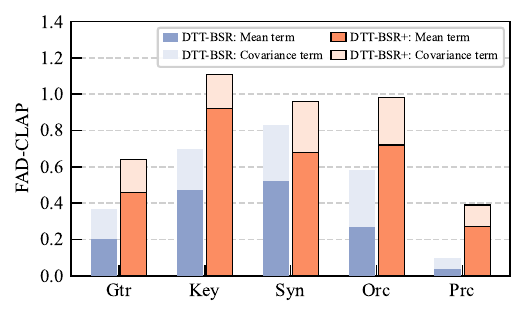}
    \vspace{-3ex}
    \caption{FAD-CLAP decomposition across five stems. Each bar represents the total FAD-CLAP score, decomposed into the mean term and covariance term.}
    \label{fig:fad_decomp}
    \vspace{-3ex}
\end{figure}

For the five stems where DTT-BSR+ yields higher FAD than DTT-BSR despite improved MMSNR, Figure~\ref{fig:fad_decomp} decomposes the FAD-CLAP into its mean term and covariance term to identify the source of this divergence. Across all five stems, the increase in total FAD is driven by the mean term, which grows substantially from DTT-BSR to DTT-BSR+, while the covariance structure of the distribution remains largely unchanged. For example, on Keyboards, the mean term increases from $0.47$ to $0.92$, whereas the covariance term decreases slightly from $0.23$ to $0.19$. This indicates that the second stage enhancement shifts the mean of the restored audio in the CLAP embedding space without altering the distributional covariance structure. These findings reveal that for certain stems, waveform-level reconstruction and semantic distribution alignment represent conflicting objectives that cannot be simultaneously optimized within a single-stage joint system. The cascade design addresses this by assigning each objective to a separate stage, achieving substantial improvements in waveform reconstruction without compromising the distribution diversity established by the previous stage.
\vspace{-1ex}
\section{Conclusion}
We present DTT-BSR+, a two-stage cascade MSR system decoupling semantic distribution fitting from signal reconstruction into separate stages, achieving consistent MMSNR improvements over the single-stage DTT-BSR across all stems. Through FAD decomposition, we reveal an implicit trade-off between signal reconstruction accuracy and semantic distribution fitting, where improving MMSNR leads to semantic mean shift while leaving distribution variance largely unchanged. The substantial performance variation across stems and the opposing FAD and MMSNR trends observed in certain stems further highlight the importance of stem-specific design in MSR systems. Our findings suggest that a unified architecture may be insufficient for all stem types, and future work should explore stem-aware approaches to optimize both reconstruction fidelity and semantic consistency.Furthermore, our future research will explore end-to-end joint training strategies to optimize global parameters, and investigate the integration of mask-free architectures, such as TF-Locoformer~\cite{10694313}, to enhance local waveform modeling and semantic consistency.

\section{Acknowledgments}
This work was supported by the National Natural Science Foundation (NSFC) of China under Grant 62471340.
The numerical calculations in this paper have been done on the supercomputing system in the Supercomputing Center of Wuhan University.

\section{Generative AI Use Disclosure}

Google Gemini was used in this paper only for polishing the manuscript.

\bibliographystyle{IEEEtran}
\bibliography{mybib}

\end{document}